\documentclass[aps,prl,twocolumn,superscriptaddress]{revtex4}

\usepackage{mathrsfs}  
\usepackage{amssymb,latexsym,amsmath}
\usepackage{graphicx}
\usepackage{braket}
\usepackage{tikz}
\usepackage{upgreek}
\usepackage{soul}

\begin{document}


\preprint{LA-UR-20-27262}

\title{Strong scattering and parallel guiding of ultracold neutrons}


\author{Zhehui Wang}
\email[]{zwang@lanl.gov}
\affiliation{Los Alamos National Laboratory, Los Alamos, NM 87545, USA}

\author{Marcel Demarteau}
\affiliation{Oak Ridge National Laboratory, Oak Ridge, TN 37830, USA}

\author{C. L. Morris}
\affiliation{Los Alamos National Laboratory, Los Alamos, NM 87545, USA}

\author{Yanhua Shih}

\affiliation{University of Maryland, Baltimore County, Baltimore, MD 21250, USA}

\date{\today}

\begin{abstract}
For ultracold neutrons with a kinetic energy below 10 neV, strong scattering, characterized by $2\pi l_{c} / \lambda\leq 1$, can be obtained in metamaterials of C and $^7$Li. Here $l_{c}$ and $\lambda$ are the coherent scattering mean free path and the neutron wavelength, respectively. UCN interferometry and high-resolution spectroscopy (nano-electronvolt to pico-electronvolt resolution) in parallel waveguide arrays of neutronic metamaterials are given as examples of new experimental possibilities. 
\end{abstract}

\pacs{}


\maketitle

With the recent advances in neutron sources such as the Spallation Neutron Source (SNS), the European Spallation Source (ESS), and the China Spallation Neutron Source (CSNS), the interdisciplinary interest in neutron science and applications continues to grow and expand globally. Due to its distinctive nuclear interaction mechanisms,  neutron scattering is indispensable to interrogation of condensed matter~\cite{Brock:1952,Shull:1953}. The thermal neutron wavelength (0.18 nm) is comparable to the lattice constant or interatomic spacing in a solid or a liquid. Unlike X-ray scattering, thermal and cold neutrons probe matter non-invasively since the neutrons essentially do not deposit energy in the sample. Neutron absorption, when $\sim$ MeV energy can be released per neutron capture, can be mitigated by nuclear isotope selection such as substituting $^1$H by $^2$H for the sample without changing the material chemistry. Meanwhile, the combination of relatively small interaction cross sections of a thermal neutron (on the order of 1 barn or 10$^{-24}$ cm$^{2}$) and the weaker neutron fluxes in comparison with a synchrotron light source, usually requires large samples ($\gg 1$ $\mu$m), as determined by the neutron scattering mean free path to the zeroth order, and long measurement time. This work is motivated by a new regime of strong neutron scattering, which has largely been unexplored to date. Strong neutron scattering forms the physics basis for neutronic metamaterials, a cousin to photonic metamaterials or photonic crystals, that can be used to enhance neutron interaction cross sections through coherent scattering and interference effects. Another purpose is to highlight new experimental possibilities enabled by neutronic metamaterials as parallel waveguides for neutron interferometry~\cite{RW:2014} and high-resolution spectroscopy in the range of nano-electronvolt (neV) to pico-electronvolt (peV) resolution, complementing neutron spin echo spectroscopy at the neV end~\cite{Mezei:2003}, and gravitational resonance spectroscopy at the peV end~\cite{JGLA:2011}.

For classical waves such as visible light, sound waves and matter waves such as electrons and cold atoms, the strong scattering regime can be reached when the Ioffe-Regel condition (also known as the Mott-Ioffe-Regel limit) is met, $2\pi l_{c} / \lambda\leq 1$~\cite{IR:1960,SZ:1986,ASB:1986,Joh:1987,KM:1993}. Here $l_{c}$ and $\lambda$ are the scattering mean free path and the wavelength of the classical waves or particles, respectively. One consequence of the strong scattering is Anderson localization~\cite{Anderson:1958}, which manifests as the complete halt of diffusion of waves as the result of scattering and interference in disordered medium. In spite of its long history and universality, direct  observation  of  Anderson localization  has been a relatively recent feat, especially for matter waves.  For example, 1D localization of cold atoms in the Bose-Einstein Condensate was reported for Rubidium-87~\cite{BJZ:2008}. 

Localization of a neutron wave or completely stopping of a neutron particle has yet to be demonstrated. Even a 3D trap with all three spatial dimensions comparable to the neutron wavelength $\lambda$ has so far been elusive. Quasibound states of neutron in matter was reported theoretically and experimentally~\cite{Kag:1970, SS:1980}. Neutron resonators for very low energy or ultracold neutrons in lattice has been considered in ~\cite{Ste:1988}. In addition to being a Fermion, neutrons provide an attractive method to study and observe localization due to a number of features such as their weak interactions with its surroundings, weak interactions amongst themselves except at extremely short distances on the order of 1 fm, finite neutron lifetime ($\sim$ 900 s) and unique beta-decay signatures, which are absent from other matter waves such as electrons or even cold atoms. Several groups have previously recognized the possibility of neutron localization in materials with neutron scattering length $a_c> 0$ or refractive index $n<1$~\cite{MFC:1998,SSMS:1998,SAB:2000}, based on the neutron total reflection, or $n \rightarrow 0$ when the neutron energy is slightly above the Fermi peudopotential for a neutron, $E  = V_F + \epsilon$. 
Observation of extremely strong neutron scattering was reported in quartz (SiO$_2$) ~\cite{MFC:1998}. An independent experimental study did not observe an onset of Anderson localization in SiO$_2$ and several other disordered media~\cite{SAB:2000}. 
For neutrons, the scattering cross section by individual atoms is small. For example, when neglecting the coherence and interference, the scattering mean free path is estimated to be 1.6 cm in graphite for density 2.26 g/cm$^3$ and neutron coherent scattering cross section 5.6$\times$10$^{-24}$ cm$^{-2}$. The neutron energy would have to be less than 8$\times$10$^{-20}$ eV to reach the Ioffe-Regel condition. Even replacing graphite by diamond powder for its larger density, neutron localization would still be impractical experimentally. Coherent scattering from multiple atoms, however, can enhance the scattering cross section and therefore reach the sufficiently large $l_{c}$ for ultracold neutrons (UCN) with kinetic energies below $\sim$ 10 neV, which is one of the main results here. 

According to Heisenberg's uncertainty principle, a neutron with a kinetic energy less than 400 neV can not exist inside a nucleus, which has a typical size in the range of 1 to 10 fm. Thermal and cold neutrons by the same token can not be contained inside a `box' comparable to a nucleus either. The neutron de Broglie wavelength ($\lambda$) is at least 50 nm for kinetic energies below 400 neV. Such cold neutrons are usually called ultracold neutrons (UCN) and can undergo total reflection at a material interface with vacuum, as first recognized by Enrico Fermi in the 1930s~\cite{GRL:1990}. The refractive index ($n$) for neutrons with a kinetic energy $E$ is $n = \sqrt{1 - V_F/E} $, with $V_F = 2\pi \hbar^2 n_a a_{c}/ m$  being the Fermi pseudopotential, $n_a$ the number density of the material, $a_{c}$ the coherent scattering length of an individual nucleus, and $m$ the neutron rest mass. Majority of the materials of stable nuclei have $a_c > 0$, and therefore $n<1$. With the carbon ($^{12}$C) $a_c = 6.65$ fm, $V_F$ is about 200 neV for graphite, 300 neV for diamond. Therefore, total reflection is only possible for UCN or neutrons with a kinetic energy below 400 neV. A few nuclei such as $^1$H, $^7$Li, $^{48}$Ti, $^{51}$V have $a_c < 0$.  In this case, neutrons would travel or be trapped inside the material made of these nuclei to undergo total internal reflection at the material-vacuum interface. UCN trapping or quasi-bound states inside a hollow C shell are also possible, which will be discussed elsewhere. With the exception of $^7$Li ($a_c$ = $-2.22$ fm,  $V_F$ = $-26.6$ neV for $n_a =4.59 \times 10^{22}$ cm$^{-3}$); however, $^1$H, $^{48}$Ti, $^{51}$V are all strong neutron absorbers, leaving $^7$Li as one of the best candidates with $a_c < 0$ and $n> 1$ for strong scattering. 

\begin{figure}[htbp] 
   \centering
   \includegraphics[width=3.5in]{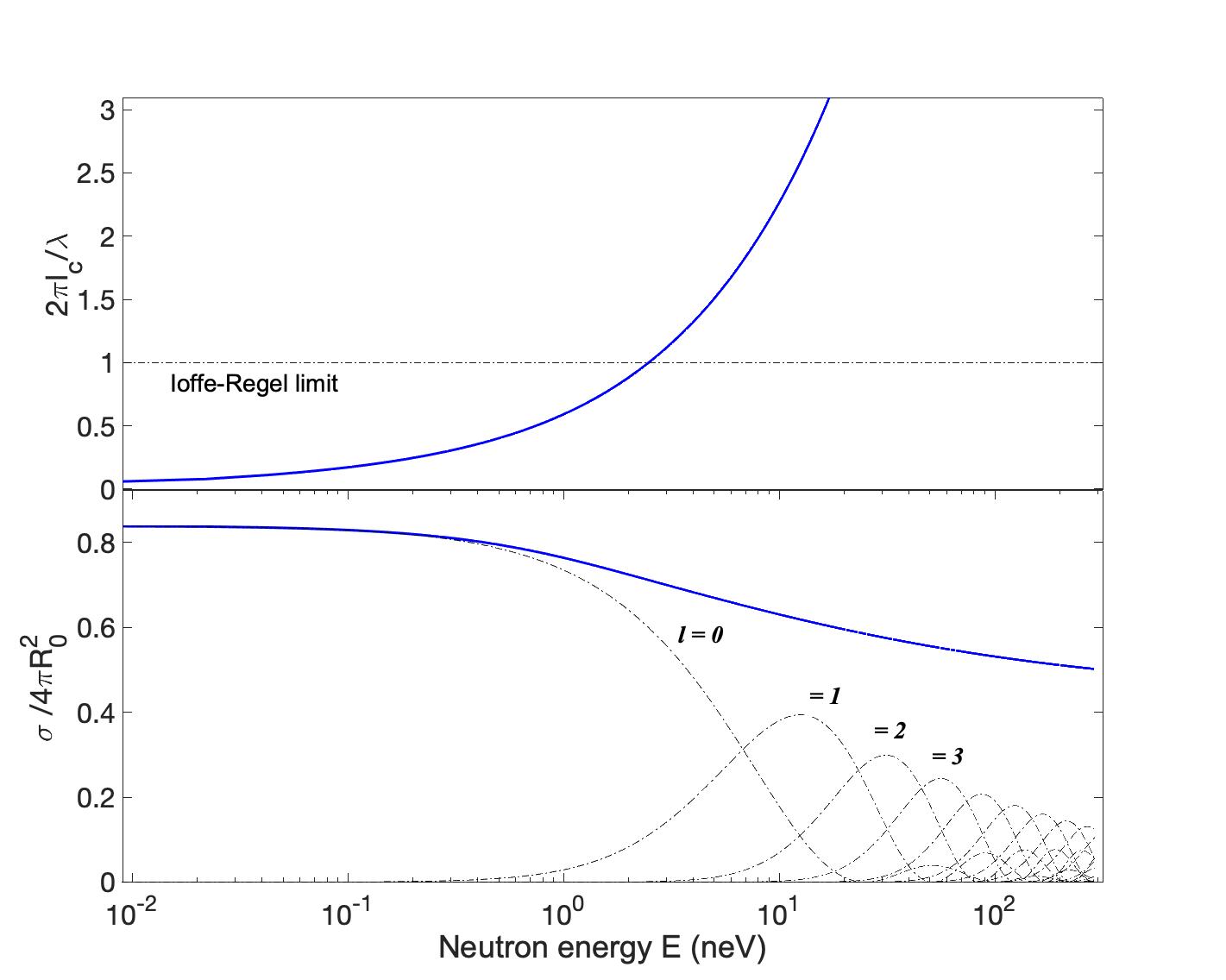}  
   \caption{Top: The normalized mean free path of the neutron according to the Ioffe-Regel condition; Bottom: UCN coherent scattering cross section (in unit of $4\pi R_0^2$) from a diamond (C)  sphere ($R_0$ = 98.7 nm). The total cross section $\sigma$ is dominated by $l=0$ for UCN energy less than 1 neV. Contributions for $l>0$ are substantial for $E>1$ neV. The strong scattering regime can only be reached for UCN energy less than 3 neV.}
   \label{fig:SScattC1}
\end{figure}

\begin{figure}[htbp] 
   \centering
   \includegraphics[width=3.5in, angle=0]{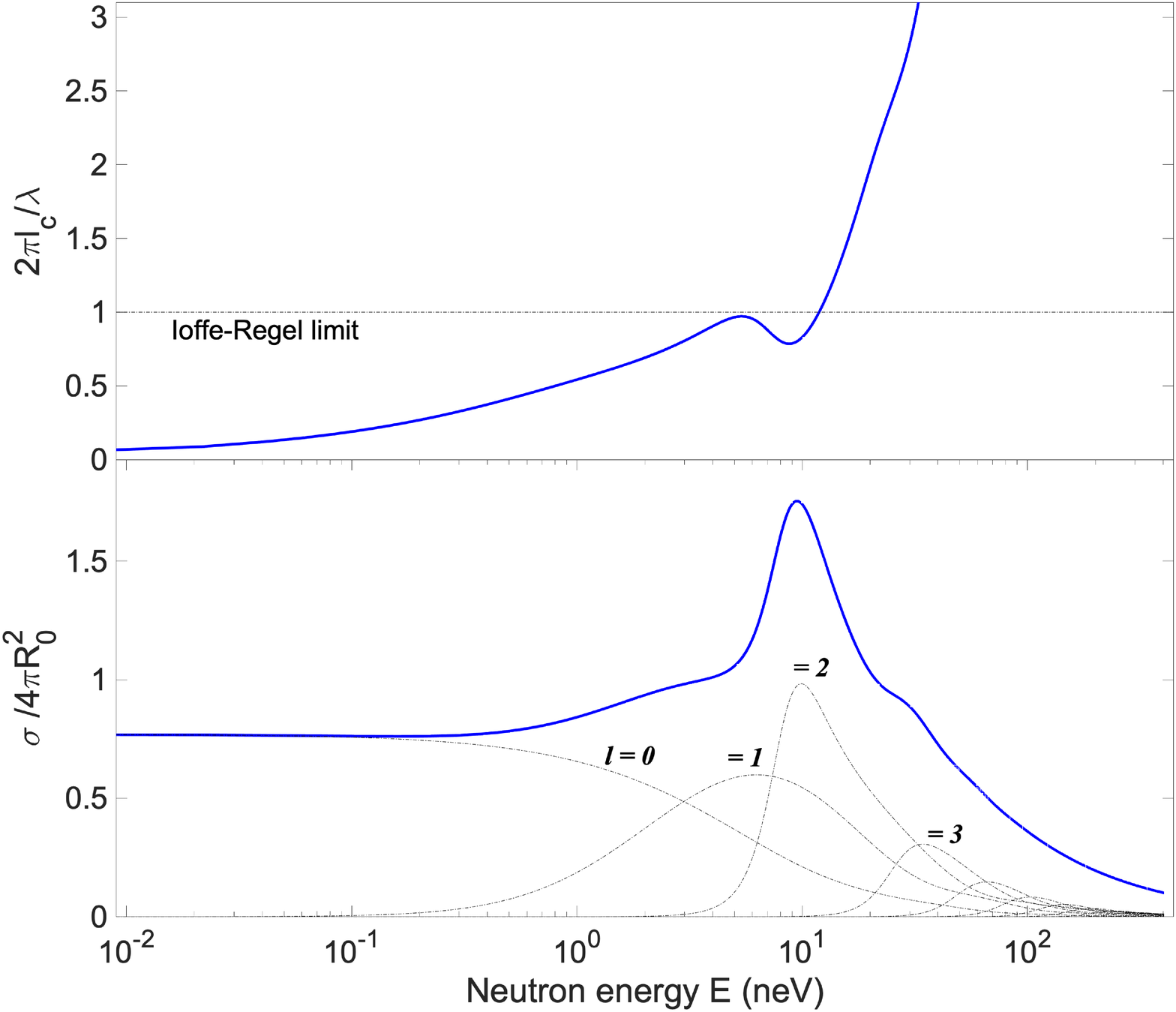} 
   \caption{Top: The normalized mean free path of the neutron according to the Ioffe-Regel condition; Bottom: UCN coherent scattering cross section (in unit of $4\pi R_0^2$) from a $^7$Li  sphere ($R_0$ = 98.7 nm). The total cross section $\sigma$ is dominated by $l=0$ for UCN energy less than 0.5 neV. Contributions to the scattering from $l>0$ partial waves are substantial for $E>1$ neV. A resonance at around 10 neV enhances the scattering cross section by more than a factor of 2 over $\sigma_0$. The strong scattering regime can be reached for $E< 13$ neV.}
   \label{fig:SScattLi1}
\end{figure}

We summarize the neutron elastic scattering cross section from small spheres of diamond or $^7$Li. The Schr\"odinger equation for a neutron has a constant scattering potential given by $V(r) =V_F$ within the radius of the sphere, $r \leq R_0$ and $V(r) = 0$ for $r > R_0$. Using the plane-wave approximation, the scattering amplitude $f(\theta)$ is given by a sum of partial waves, each characterized by its angular momentum $l$ and a corresponding phase shift $\delta_l$~\cite{MM:1933}
\begin{equation}
f(\theta) = \frac{1}{k} \sum_{l=0}^\infty (2l+1) e^{i\delta_l} \sin \delta_l P_l (\cos \theta),
\end{equation}
with $\hbar k = \sqrt{2mE}$ being the incident neutron momentum. The differential scattering cross section is $d\sigma/d\Omega = |f(\theta)|^2$. The total scattering cross section is
\begin{equation}
\sigma =\frac{4 \pi}{k^2} \sum_{l=0}^\infty (2l+1) \sin^2\delta_l,
\end{equation}
where the phases $\delta_l$ are obtained from the solution of the equation
\begin{equation}
\frac{d^2 \chi_l}{dr^2} + \left[k^2 - \frac{2 m}{\hbar^2} V(r) - \frac{l(l+1)}{r^2}\right] \chi_l =0
\label{eq:sch1x}
\end{equation}
with the asymptotic approximation when $kr \rightarrow \infty$, $\chi_l \sim \sin (kr - \frac{1}{2} l\pi + \delta_l)$. The usual boundary conditions of continuity for $\chi_l$ and its first-order derivative $d\chi_l/dr $ apply at the sphere surface $r=R_0$. The calculated cross sections for C and $^7$Li, together which the scattering mean free path $l_c$, are shown as a function of neutron energy in Figs.~\ref{fig:SScattC1} and \ref{fig:SScattLi1}, in which the scattering mean free path is given by $l_c = 4\pi R_0^3/3f_p \sigma $ for the the packing fraction $f_p$ of the spheres. Constrained by geometry, $f_p \le 0.74 $. We use $f_p = 0.5$ here in both figures.

In the low energy limit ($kR_0, E \rightarrow 0$), both cross sections are dominated by the S-wave ($l=0$) scattering as expected. For C, $\displaystyle{\sigma \sim \sigma_0 =
4 \pi R_0^2 ( 1 - \frac{\tanh \kappa_+ R_0}{\kappa_+ R_0})^2}$, 
with $\displaystyle{\kappa_+^{2} = \frac{8\pi^2 m}{h^2}V_F}$. For $^7$Li, $\displaystyle{\sigma \sim \sigma_0 = 4\pi R_0^2 (\frac{\tan \kappa_- R_0}{\kappa_- R_0} - 1)^2}$ with $\displaystyle{\kappa_-^2 = - \frac{8\pi^2 m}{h^2}V_F}$.  A resonance scattering peak is found for the $^7$Li sphere when $E = 10$ neV. The resonance energy is consistent with the a bound-state solution to the Schr\"odinger equation for the $^7$Li sphere. The minimum sphere radius $r_c$ for the existence of a 3D neutron bound state in $^7$Li is 
\begin{equation}
R_0 \geq r_c = \frac{1}{4}\sqrt{\frac{\pi}{n_aa}}.
\end{equation}
$r_c$= 44 nm at the solid $^7$Li density $n_a$ = 4.63 $\times$ 10$^{22}$ cm$^{-3}$. Calculations of the $\sigma_l/\sigma$ as a function of $^7$Li radius confirm that the contributions from non-zero angular momenta $l\geq 1$ to the scattering cross section are negligible for $R_0 < $ 50 nm.

Both strong scattering scenarios when the condition $2\pi l_{c} / \lambda\leq 1$ is met, whether it is resonance-enhanced (as in $^7$Li) or not (as in C), are of significant experimental interest. Not only because both predictions are new, but more interestingly, a 3D bound state of a UCN on a spatial scale when the matter wave properties of a neutron matter has not been achieved in the laboratory.  The recent progress in this new quantum regime has already led to fruitful experimental results such as the first observation of quantum states of neutrons in the Earth's gravitational field~\cite{Nes:2002} or gravitational resonance spectroscopy~\cite{JGLA:2011}. However, such successes have been relying on `flying' neutrons, when the quantum effects are only significant in 1D. Both Anderson localization and the bound states of neutrons centered at a $^7$Li sphere offer the possibility of a new type of 3D `neutron quantum bottle', when the matter wave properties manifest in 3D. Such 3D neutron waves are more sensitive quantum probes of physics beyond the Standard Model than the existing 1D case in several ways~\cite{Abele:2010, Snow:2018}. One advantage is through longer interaction time for higher energy resolution, $\Delta E = \hbar \Delta \Phi /T$. The Ramsey's method for flying neutrons as described in~\cite{Abele:2010} corresponds to $T \sim 100$ ms and $\Delta E = 5 \times 10^{-17}$ eV. The phase accuracy $\Delta \Phi$ is limited by the neutron counting statistics and UCN source intensity to about 10$^{-2}$ radians. For a UCN confined to a 3D quantum bottle, quantum nondemolition (QND) measurement of a neutron may be possible through probing the neutron magnetic momentum and avoiding UCN capture by a nucleus such as $^3$He, $^6$Li and $^{10}$B. In the QND measurement scenario, $\Delta \Phi$ may be enhanced by another factor $ \sim 10^3$, which is the number of times that the neutron wave function is measured repetitively without destruction. We assume that the neutron wavefunction is measured at 1 Hz for the duration of the neutron lifetime. Since the magnetic field due to neutron nuclear momentum decays with the distance from the neutron ($r$) as $1/r^3$, it is advantageous that a UCN is localized to a volume comparable to $\lambda^3$ throughout the QND measurement.

{\it Scattering experiments} The scattering rate from an individual sphere is given by   $\displaystyle{\frac{1}{4}n_0 \langle \sigma v \rangle} = \frac{1}{4}n_0 \int dE \sigma(E) \sqrt{\frac{2E}{m}} f(E) $ for a UCN density of $n_0$ per unit volume and the energy distribution function $f(E)$. The normalization condition is $\displaystyle{\int dE f(E)} = 1$. For a Maxwell-Boltzman energy distribution at a temperature $k_BT$~\cite{GRL:1990} and total cross section as given in Fig.~\ref{fig:SScattLi1}, the scattering rate per sphere peaks at 5.8$\times$10$^{-6}$ Hz for the UCN density of 10$^2$ cm$^{-3}$ and UCN temperature about 20 neV. The scattering rates are 4.6$\times 10^{-6}$, 3.6$\times 10^{-6}$ and 2.2$\times 10^{-6}$ Hz for neutron temperatures 100, 200 and 500 neV respectively. The scattering rate decreases roughly as $E^\alpha$, with $\alpha = -0.39 \pm 0.13$ for $ 80 < k_BT < 500$ neV.   It is therefore feasible to use $10^6$ nanospheres in parallel to achieve a scattering rate of a few Hz for the existing UCN sources. An area of 10$^2 \times 10^2$ $\upmu$m$^2$ will be suffice. 

When a large number of nanospheres are used, the corresponding total scattering cross sections  are modulated from that of a single sphere by a multiplier known as the structure factor $S({\bf k})$~\cite{GRL:1990}. Additional modifications to the cross section come from phonon scattering and incoherent scattering~\cite{Ste:1980}. To reduce the complexity of the scattering and interference effects to a level that is more conducive to theory and especially analytic theory, we may carry out the scattering experiments with the aid of waveguides for ultracold neutrons, when the scattering objects are be placed at the end of the waveguides. As given below, guided UCN modes have quantized energy levels. Correspondingly, the scattering cross section as given above for Maxwell-Boltzmann distribution $\displaystyle{\int dE f(E)}$ can be readily modified to a sum over discrete energy levels $\displaystyle{\sum_{k_x, n_y, E_n } p(k_x, n_y, E_n)}$ as given in Eq.~(\ref{eq:modes}) for the mode probability $p(k_x, n_y, E_n)$. We shall also mention that, for a sufficiently disordered arrangement of scattering spheres, we may also experimentally search for neutron Anderson localization. Experiments with guided neutrons will reduce the search from 3D disordered objects to 1D and 2D systems, where analytical results have shown a localization length scale on the order of $l_c$ (1D) and  $l_c \exp(k l_c)$ (2D)~\cite{Joh:1987,KM:1993}.

{\it Quantized modes in a neutron waveguide} Neutron guides, originally proposed by Maier-Leibnitz
and demonstrated in 1962 by Christ and Springer, allow
the transport of neutron beams over relatively long distances
with small and sometimes negligible loss in phase-space density. Different ways of neutron guiding, mostly for thermal neutrons, have since been studied by a number of authors~\cite{CDM:1992,Dab:2003,CA:2004}. The wave properties of neutrons in these guides can often be ignored, corresponding to the `ray optics' regime in light transportation. Exceptions do exist, as demonstrated for example in the well-known Colella, Overhauser, and  Werner (COW) experiment in 1975, which used an exquisite design of monolithic silicon wafer for neutron guiding, and beam splitting through diffraction. 

\begin{figure}[htbp] 
   \centering
   \includegraphics[width=3.2in,angle=0]{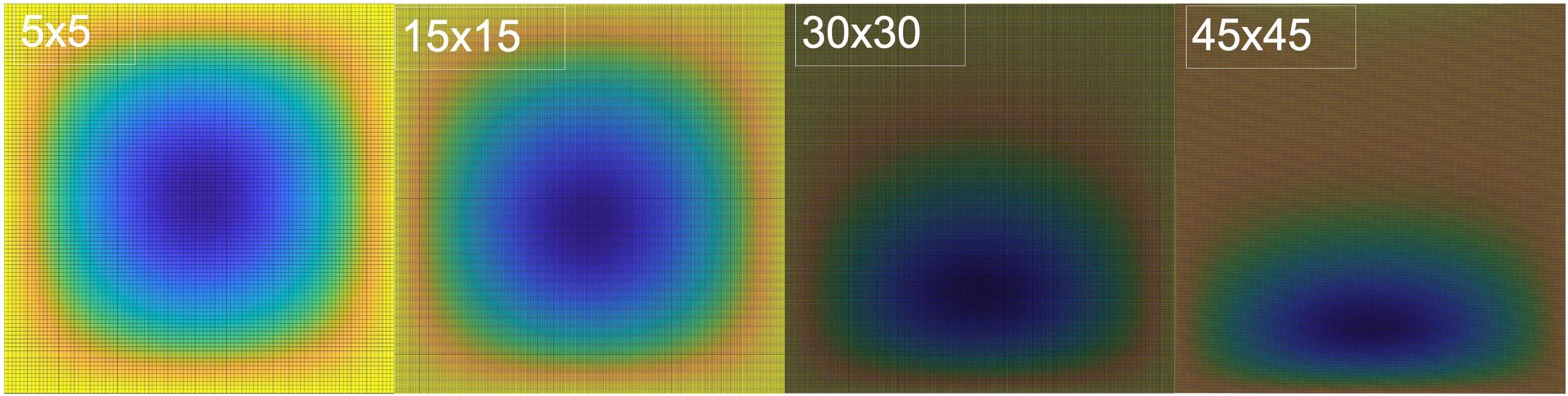} 
   \caption{Neutron distribution probability contours inside square waveguides for the lowest mode number $n_y =1$ in the horizontal direction and $E_1$ in the vertical direction (along the Earth's gravity). $L_y = L_z$ = 5, 15, 30 and 45 $\upmu$m from left to right. The asymmetric distributions in the vertical direction are due to the quantized gravitational states of the neutron.}
   \label{fig:mode1}
\end{figure}

For ultracold neutrons, we have to take into account of the wave properties of neutrons and use Schr\"odinger equation for the transport. In contrast to photon propagation in light waveguides, the gravity effects can no longer be ignored; therefore, the orientation of the waveguide relative to the Earth's gravity affects the propagation. We consider the neutron eigenmodes in a horizontally positioned rectangular waveguide along the $x$-axis with the cross section area given by $L_y \times L_z$.  The $z$-axis is along the direction of the Earth's gravity. By choosing the coordinate origin at one of the lower corners of the wave guide, the eigenmodes of neutron wave function (to within a normalization factor) are given by
\begin{equation}
\begin{split}
&\psi (k_x, n_y, E_n) = e^{ik_xx} \sin (\pi n_y \frac{y}{L_y})  \\
& \times \left[{\rm Bi}(\tilde{z}_0) {\rm Ai} (\frac{z}{l_0} - \frac{ E_n}{E_0}) -{\rm Ai}(\tilde{z}_0){\rm Bi}(\frac{z}{l_0} - \frac{E_n}{E_0}) \right],  
\end{split}
\label{eq:modes}
\end{equation}
where $k_x$ is a positive continuous number for propagating modes along the $x$-direction. $n_y $ is a positive integer that characterizes the modes in the horizontal direction. $E_n$ is the discrete energy levels in the direction of the gravity. Ai$(x)$ and Bi$(x)$ are Airy functions of the first and second kind. The characteristic length $\displaystyle{l_0 = (\frac{\hbar^2}{2m^2 g})^{1/3}}$ = 5.87 $\upmu$m for a neutron in the Earth's gravitational acceleration $g$. The characteristic gravitational energy is $E_0 = mgl_0 = 0.602$ peV. A few examples of the lowest mode-number eigenfunctions are shown as a function of the size of square waveguides in Fig.~\ref{fig:mode1}.
The gravitational effect clearly shows as the asymmetry of the neutron probability distribution towards the bottom of the waveguide as the vertical dimension $L_z$ increases from 5 $\upmu$m (comparable to $l_0$) to 45 $\upmu$m. Similar trend is also observable for higher mode number $n$ in the direction of the Earth's gravity. 

\begin{figure}[htbp] 
   \centering
   \includegraphics[width=3.5in,angle=0]{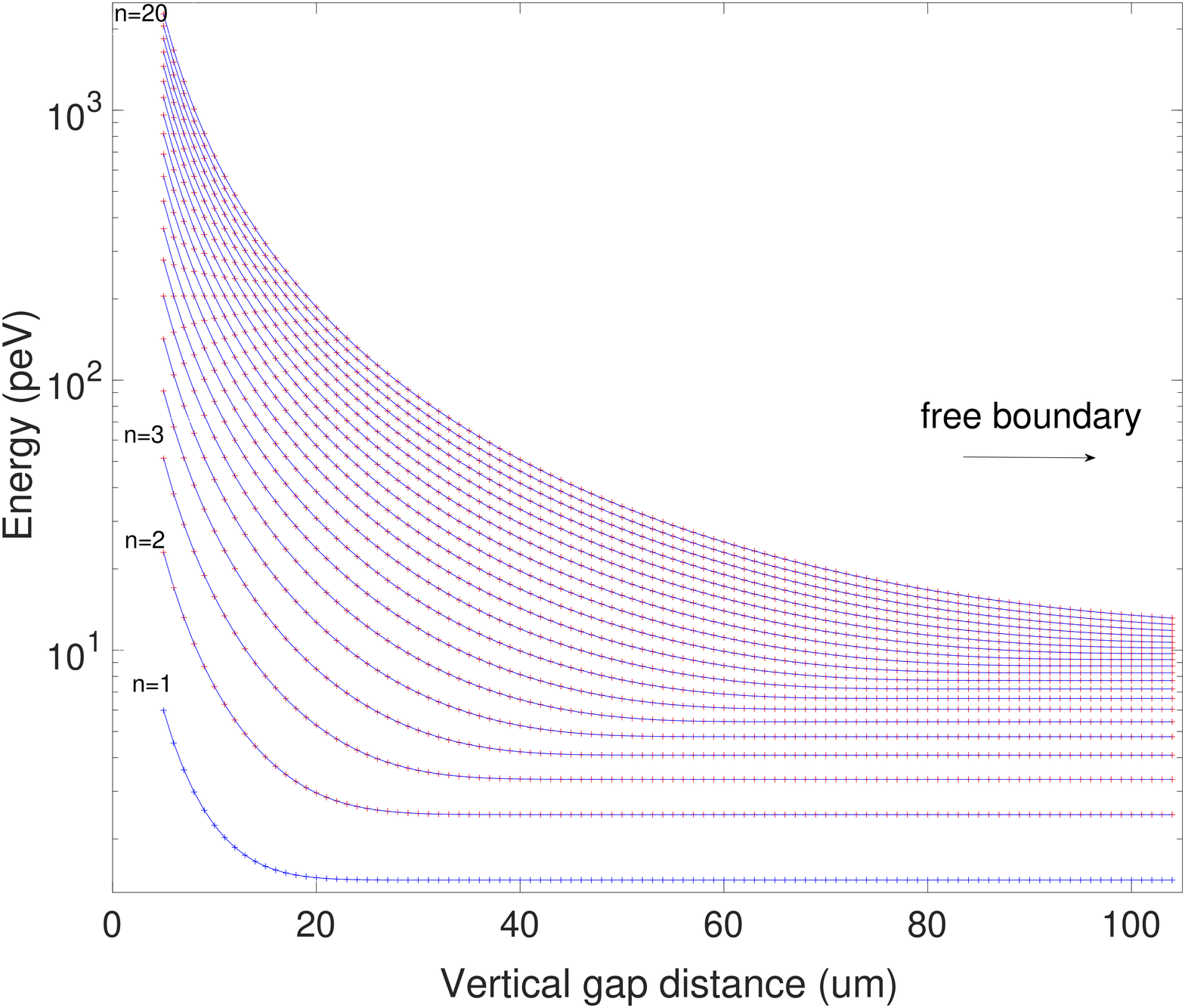} 
   \caption{Quantized energy levels $E_n$ (up to mode number $n=20$) of ultracold neutrons inside a waveguide, which are modulated by gravity, as a function of the waveguide vertical gap distance $L_z$. $E_n$ approach the values of a free boundary when the $L_z$ is large compared with $l_0$.}
   \label{fig:quantized1}
\end{figure}

The vertical energy levels $E_n$ are obtained from
\begin{equation}
 \left|\begin{array}{cc} {\rm Ai}(\tilde{z}_0)&  {\rm Bi}(\tilde{z}_0)\\  {\rm Ai}(\tilde{z}_1)&  {\rm Bi}(\tilde{z}_1)  \end{array}\right| = 0,
 \label{eq:Type2Zeros1}
\end{equation}
with $\tilde{z}_0 =  - E_n/E_0$ and $\tilde{z}_1 = L_z/l_0 - E_n/E_0$. The $E_n$ values for $n$ up to 20 are shown in Fig.~\ref{fig:quantized1} as a function of the gap distance. While the energy levels in the horizontal direction is quadratically spaced with respect to $n_y^2$ as $\displaystyle{\frac{ \pi^2 \hbar^2n_y^2}{2m L_y^2}}$, the energy levels in the vertical direction $E_n$ depend on both the waveguide dimension $L_z$ and the mode number $n$ in a more complex way related to the Airy functions. The following Bohr-Sommerfeld formula is sometimes used as the approximation, $\displaystyle{E_n= E_0 \left[ \frac{3\pi}{2} (n -\frac{1}{4})\right]^{2/3}}$, when $L_z \gg l_0$ (or the free boundary condition). If $1/4$ is replaced by $63/256$, the approximation improves except for $E_1$.

We mention that the quantized energy levels and their susceptibility to gravity make the UCN gravitational spectroscopy possible in waveguides, as previously demonstrated for the free boundary configuration~\cite{JGLA:2011}. For material applications, UCN waveguide spectroscopy could be useful for the neutron energy transition ($\Delta E$) induced by materials in range of 1 $\upmu$eV (waveguide dimension $L_y \sim$ 10 nm) to $ \sim 1 $ peV ($L_y \sim$ 10 $\upmu$m), bridging the energy resolution gap between neutron spin echo spectroscopy~\cite{Mezei:2003} and gravitational resonance spectroscopy.

\begin{figure}[htbp] 
   \centering
   \includegraphics[width=3in]{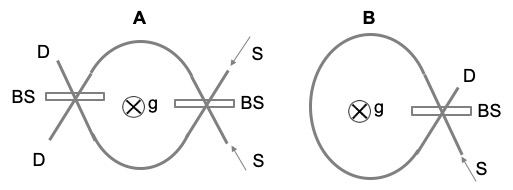}  
   \caption{{\bf A}. Mach-Zehnder interferometer configuration based on UCN waveguide. {\bf B}. Ring interferometer configuration. S: UCN source; D: UCN Detector; BS: Beam splitter; The direction of the gravity (g) is perpendicular to the plane of the beam paths to the zeroth order.}
   \label{fig:EXPInt1}
\end{figure}

{\it UCN interferometers in waveguides}  
Similar to light interferometers using optical fiber waveguides~\cite{Chow:1985, CSP:2009}, we may consider UCN interferometry in waveguides. A variety of interferometer configurations are possible~\cite{RW:2014}. We only examine the Mach-Zehnder interferometer (MZI) setup in a waveguide, Fig.~\ref{fig:EXPInt1}A. A slight modification to the MZI configuration can lead to a ring interferometer, Fig.~\ref{fig:EXPInt1}B. 

Compared with the Colella, Overhauser, and  Werner (COW) experiment for thermal neutrons, MZI setup as shown in Fig.~\ref{fig:EXPInt1} is able to remove the additional BS in-between the source S and the detector D by the waveguide. As shown in~\cite{Chow:1985, CSP:2009}, the total phase-shift is given by a sum of three main contributions, $\displaystyle{\Delta\Phi =\Delta \Phi_{COW}+\Delta \Phi_{Sagnac} + \Delta \Phi_{mat.}}$. Here  the COW phase due to gravity is $\displaystyle{\Delta \Phi_{COW} = \lambda \frac{m_im_g}{2\pi\hbar^2} g A_0 \cos \alpha} $, the Sagnac phase due to the Earth's rotation $\displaystyle{\Delta \Phi_{Sagnac} = \frac{2m_i}{\hbar} \boldsymbol{\Omega}\cdot{\bf A}_0}$, and materials effects at the BS, $\Delta \Phi_{mat.} = q_{mat.} \cos \alpha$. When the angle $\alpha = \pi/2$, or the plane of the interferometer is perfectly perpendicular to the direction of the gravity, $\Delta\Phi$ has no explicit dependence on the UCN wavelength $\lambda = 2\pi/k_x$ as defined in Eq.~(\ref{eq:modes}) for each waveguide mode, making the interferometer broadband that is only sensitive to $\Delta \Phi_{Sagnac}$. Meanwhile, the COW phase $\Delta \Phi_{COW}$ can be useful to study the equivalence principle and general relativity in this new neutron energy regime. A UCN waveguide interferometer is potentially more sensitive by a factor of 10$^3$ to 10$^4$ due to the longer neutron wavelength $\lambda$, the trade-off is that the UCN flux is much weaker. Longer experimental time and using an array of interferometers in parallel configuration to intercept as much as UCN flux are two possible remedies. 

In summary, we show that for neutrons with a kinetic energy below 10 neV, strong scattering can be obtained in metamaterials of C and $^7$Li  with a feature size on the order of 100 nm. Such strong scattering forms the basis for UCN waveguiding in metamaterials, and allow UCN interferometry and high-energy resolution spectroscopy in the energy range of pico-electronvolt to nano-electronvolt. Parallel waveguide arrays of neutronic metamaterials, when micro-fabricated with other neutron optics components such as beam splitters, and neutron detectors, are attractive options to boost the experiment data rates for the currently available UCN fluxes. 

Z.W. wishes to thank Prof.~Tongcang Li (Purdue University), Dr. Stephan Eidenbenz (LANL) for stimulating discussions. This work is supported by the LANL LDRD program.

\end{document}